\newcommand*{\ATLASLATEXPATH}{latex/}
\documentclass[UKenglish,texlive=2011,cernpreprint,block=space]{\ATLASLATEXPATH atlasdoc}
\pdfoutput=1
\usepackage{\ATLASLATEXPATH atlaspackage}
\usepackage{\ATLASLATEXPATH atlasbiblatex}

\usepackage{\ATLASLATEXPATH atlascontribute}

\usepackage{\ATLASLATEXPATH atlasphysics}
\graphicspath{{logos/}{figures/}}

\usepackage{TOPQ_2015_01-defs}

\newif\ifauxmat
\auxmatfalse

\AtlasTitle{Evidence for single top-quark production in the
$\boldsymbol{s}$-channel in proton--proton collisions at
$\boldsymbol{\sqrt{s}\!=\!\unit[8]{\TeV}}$ with the \ATLAS detector using the
Matrix Element Method}

\author{The ATLAS Collaboration}

\date{\today}

\PreprintIdNumber{CERN-PH-EP-2015-286}

\AtlasDate{\today}

\AtlasJournal{PLB}
\AtlasAbstract{%
This Letter presents evidence for single top-quark
production in the $s$-channel using proton--proton collisions at a
centre-of-mass energy of $\unit[8]{\TeV}$ with the \ATLAS detector at
the \CERN Large Hadron Collider. The analysis is performed on events
containing one isolated electron or muon, large missing transverse
momentum and exactly two $\bquark$-tagged jets in the final state. The
analysed data set corresponds to an integrated luminosity of
$\unit[20.3]{fb^{-1}}$. The signal is extracted using a
maximum-likelihood fit of a discriminant which is based on the matrix
element method and optimized in order to separate single-top-quark
$s$-channel events from the main background contributions, which are
top-quark pair production and $\Wboson$ boson production in
association with heavy-flavour jets. The measurement leads to an
observed signal significance of 3.2 standard deviations and a measured
cross-section of
$\sigma_s\!=\!\unit[4.8\!\pm\!0.8\text{(stat.)}^{+1.6}_{-1.3}\text{(syst.)}]{pb}$,
which is consistent with the Standard Model expectation. The expected
significance for the analysis is 3.9 standard deviations. }

\AtlasCoverSupportingNote{}
\hypersetup{pdftitle={Evidence for single top-quark production in the
    s-channel in proton-proton collisions at sqrt(s)=8TeV with the
    ATLAS detector using the Matrix Element Method},pdfauthor={The
    ATLAS Collaboration}}

\begin{document}
\nolinenumbers

\maketitle

%\tableofcontents

% List of contributors - print here or after the Bibliography.
%\PrintAtlasContribute{0.30}
%\clearpage

%\begin{multicols}{2}
%\multicollinenumbers

%-------------------------------------------------------------------------------

\section{Introduction}
\label{sec:intro}

\begin{figure}[hb]
  \centering
  \includegraphics[width=0.26\textwidth]{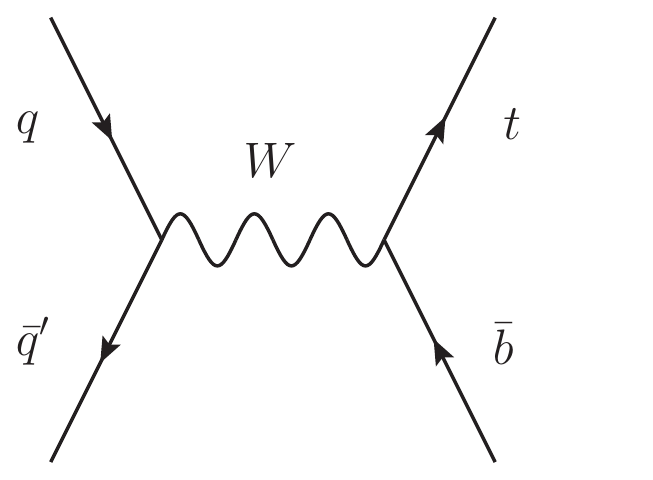}
  \caption{Feynman diagram in leading-order QCD for the dominant hard
    scattering process in the $s$-channel of single top-quark
    production.}
  \label{fig:feynman}
\end{figure}

In proton--proton\ ($\pp$) collisions, top quarks are produced mainly
in pairs via the strong interaction, but also singly via the
electroweak interaction through a $\Wtb$ vertex. Therefore, single
top-quark production provides a powerful probe for the electroweak
couplings of the top quark. In the Standard Model (SM), three
different production mechanisms are possible in leading-order\ (LO)
QCD: an exchange of a virtual $\Wboson$ boson either in the
$t$-channel or in the $s$-channel (see Fig.~\ref{fig:feynman}), or the
associated production of a top quark and a $\Wboson$ boson. Among
other interesting features, $s$-channel single top-quark production is
sensitive to new particles proposed in several models of physics
beyond the SM, such as charged Higgs boson or $\Wprime$ boson
production\ \cite{Tait:2000sh}. It\ also plays an important role in
indirect searches for new phenomena that could be modelled as
anomalous couplings in an effective quantum field theory\
\cite{Cao:2007ea}. Furthermore, $s$-channel production, like the other
two production channels, provides a direct determination of the
absolute value of the Cabibbo--Kobayashi--Maskawa (CKM) matrix element
$V_{\topquark\bquark}$.

Single top-quark production was first seen by the \CDF and \Dzero
collaborations in combined measurements of the $s$-channel and
$t$-channel\ \cite{Aaltonen:2009jj,Abazov:2009ii}. Recently, the
$s$-channel alone was observed in a combination of the results from
both collaborations\ \cite{CDF:2014uma}. At the Large Hadron Collider
(\LHC)\ \cite{lhc} the production of single top quarks was observed
both in the $t$-channel and in associated $\Wt$ production by the
\CMS\ \cite{Chatrchyan:2012ep,PhysRevLett.110.022003} and \ATLAS
collaborations\ \cite{Aad:2012ux,Aad:2012dj}. For the $s$-channel,
results of a search at $\rts\!=\!\unit[8]{\TeV}$ using an integrated
luminosity of $\unit[20.3]{\ifb}$\ were published by \ATLAS\
\cite{Aad:2014aia}. That analysis was based on a boosted decision tree
(BDT) event classifier and led to an upper limit of $\unit[14.6]{pb}$
at the 95\% confidence level. The obtained cross-section was
$\sigma_{s}^{\text{BDT}}\!=\!\unit[5.0\pm4.3]{pb}$ with an observed
signal significance of 1.3\,$\sigma$.

Standard Model predictions are available for the production of single
top quarks in next-to-leading-order\ (NLO) QCD\
\cite{Bordes:1994ki,Stelzer:1998ni,Stelzer:1997ns} including resummed
next-to-next-to-leading logarithmic\ (NNLL) corrections for soft gluon
emissions \ \cite{Kidonakis:2011wy,Kidonakis:2010tc,Kidonakis:2010ux}.
For the $s$-channel the predicted total inclusive cross-section for
$\pp$ collisions at a centre-of-mass energy
$\sqrt{s}\!=\!\unit[8]{\TeV}$ is
$\sigma^{\text{th}}_s\!=\!\unit[5.61\!\pm\!0.22]{pb}$, while for the
$t$-channel it\ is
$\sigma^{\text{th}}_t\!=\!\unit[87.76^{+3.44}_{-1.91}]{pb}$, and
$\sigma^{\text{th}}_{\Wt}\!=\!\unit[22.37\!\pm\!1.52]{pb}$ for
associated $\Wt$ production. The given uncertainties include
variations of the renormalization and factorization scales, as well as
an estimate of the uncertainty of the parton distribution function
(PDF) needed for the calculation.

In this Letter, a measurement of single top-quark $s$-channel
production in $\pp$ collisions with $\sqrt{s}\!=\!\unit[8]{\TeV}$ at
the \LHC is presented. Each of the two other single-top-quark
production processes, $t$-channel and $\Wt$ production, is treated as
a background process assuming its cross-section as predicted by
NLO+NNLL QCD calculations. In the SM the top quark decays almost
exclusively into a $\Wboson$ boson and a $\bquark$-quark. This
analysis considers only the leptonic decays ($\electron$ or $\muon$)
of the $\Wboson$ boson, since the fully hadronic final states are
dominated by overwhelming multi-jet background. Some of the events
containing a $\Wboson$ boson decaying into a $\tau$ lepton which
subsequently decays leptonically are also selected. At LO the final
state contains two jets with large transverse momenta: one jet
originating from the decay of the top quark into a $\bquark$-quark
(``$\bquark$-jet''), and another $\bquark$-jet from the $\Wtb$ vertex
producing the top quark. Thus the experimental signature consists of
an isolated electron or muon, large missing transverse momentum,
$\Etmiss$, due to the undetected neutrino from the $\Wboson$ boson
decay, and two jets with large transverse momentum, $\pT$, and which
are both identified as containing $\bquark$-hadrons
(``$\bquark$-tagged''). The electron and muon channels in this
analysis are merged regardless of the lepton charge in order to
measure the combined production cross-section of top quarks and top
antiquarks.

In contrast to the aforementioned BDT-based analysis\
\cite{Aad:2014aia}, the signal extraction in this analysis is based on
the matrix element (ME) method\ \cite{Kondo:1988yd,Kondo:1991dw}. The
same data set is used in both analyses. This analysis takes advantage
of enhanced simulation samples which reduce the statistical
uncertainty and give a better description of the data. Furthermore,
updated calibrations for the 2012\ data are used, resulting in a
reduction of systematic uncertainties. The event selection is improved
by adding a veto on dileptonic events, which leads to a significant
suppression of the background for top-quark pair ($\ttbar$) production
(see Section~\ref{sec:selection}). The combination of all these
measures results in a significant improvement in the sensitivity to
the $s$-channel process. Approximately half of this improvement can be
attributed to the change in method from BDT to ME. In particular, the
BDT technique applied to this analysis is limited by the sample sizes
available for the training, while the ME approach is not sensitive to
this limitation.

%-------------------------------------------------------------------------------

\section{The \ATLAS detector}
\label{sec:detector}

The \ATLAS detector\ \cite{ATL-2008-001} is a multi-purpose detector
consisting of a tracking system, calorimeters and an outer muon
spectrometer. The inner tracking system contains a silicon pixel
detector, a silicon microstrip tracker and a straw-tube transition
radiation tracker. The system is surrounded by a thin solenoid magnet
which produces a $\unit[2]{T}$ axial magnetic field, and it provides
charged-particle tracking as well as particle identification in the
pseudorapidity\footnote{\ATLAS uses a right-handed coordinate system
  with its origin at the nominal interaction point (IP) in the centre
  of the detector and the $z$-axis along the beam pipe. The $x$-axis
  points from the IP to the centre of the \LHC ring, and the $y$-axis
  points upward. Cylindrical coordinates $(r,\phi)$ are used in the
  transverse plane, $\phi$ being the azimuthal angle around the beam
  pipe. The pseudorapidity is defined in terms of the polar angle
  $\theta$ as $\eta\!=\!-\ln\tan(\theta/2)$.} region $|\eta|\!<\!2.5$.
The central calorimeter system covers the range of $|\eta|\!<\!1.7$
and is divided into a liquid-argon electromagnetic sampling
calorimeter with high granularity and a hadron calorimeter consisting
of steel/scintillator tiles. The endcap regions are equipped with
liquid-argon calorimeters for electromagnetic and hadronic energy
measurements up to $|\eta|\!=\!4.9$. The outer muon spectrometer is
immersed in a toroidal magnetic field provided by air-core
superconducting magnets and comprises tracking chambers for precise
muon momentum measurements up to $\abseta\!=\!2.7$ and trigger
chambers covering the range $\abseta\!<\!2.4$. The combination of all
these systems provides efficient and precise reconstruction of leptons
and photons in the range $|\eta|\!<\!2.5$. Jets and $\Etmiss$ are
reconstructed using energy deposits over the full coverage of the
calorimeters, $|\eta|\!<\!4.9$. A three-level trigger system\
\cite{Aad:2012xs} is used to reduce the recorded rate of uninteresting
events and to select the events in question.

%-------------------------------------------------------------------------------

\section{Data and simulation samples}
\label{sec:samples}

The data for this analysis was collected with the \ATLAS detector
at the \LHC in 2012 at a centre-of-mass energy of $\unit[8]{\TeV}$
using single-electron or single-muon triggers. The applied trigger
thresholds ensure a constant efficiency with respect to the energy of
the lepton candidates used in this analysis. Each triggered event
includes on average about 20 additional $\pp$ collisions (pile-up)
from the same bunch-crossing. Only events recorded under stable beam
conditions are selected and all events have to pass stringent data
quality requirements and need to contain at least one reconstructed
primary vertex with at least five associated tracks. The data used by
this analysis corresponds to an integrated luminosity of
$\unit[20.3\!\pm\!0.6]{fb^{-1}}$\ \cite{Aad:2013ucp}.

The samples used for the simulation of the single-top-quark
$s$-channel signal events, as well as the ones for the $\ttbar$,
single-top-quark $t$-channel and $\Wt$ backgrounds, were produced
using the NLO generator \PowhegBox\ (v1\_r2129) \cite{Frixione:2007vw}
with the CT10\ PDFs\ \cite{Lai:2010vv}. The parton shower,
ha\-dronization and underlying event were simulated with \Pythia\
(v6.42)\ \cite{Sjostrand:2006za} using the Perugia\ 2011C set of tuned
parameters\ \cite{Skands:2010ak}. For generator, parton shower and
fragmentation modelling studies, alternative simulation samples are
employed. In case of the $s$-channel signal and the $t$-channel
background the \aMCatNLO\ (v2.0) generator was used\
\cite{Alwall:2014hca}, while for the $\ttbar$ and $\Wt$ backgrounds
it\ was the \MCatNLO\ (v4.03)\ \cite{Frixione:2005vw} generator. In
both cases the CT10\ PDFs were used and the generators were interfaced
to \Herwig\ (v6.52)\ \cite{Corcella:2000bw} for parton showering and
hadronization, and \Jimmy\ (v4.31)\ \cite{Butterworth:1996zw} for the
underlying event. The impacts of scale variations as well as
uncertainties on the initial and final state radiation (ISR/FSR) in
signal events were studied using samples generated with the \PowhegBox
generator, again connected to \Pythia, for various values of the
factorization and renormalization scales.

The processes for $\Wboson$ boson production in association with jets
($\Wjets$) were modelled by the LO mul\-ti-parton \Sherpa generator\
(v1.4.1)\ \cite{Gleisberg:2008ta} together with CT10\ PDF sets. This
generator matches the parton shower to the multi-leg LO matrix
elements by using the CKKW method\ \cite{Catani:2001cc}. The \Sherpa
generator was used for the complete event generation including the
underlying event, using the default set of tuned parameters. The
background contributions from $\Zboson$ boson production in
association with jets ($\Zjets$) were simulated using the LO\ \Alpgen\
(v2.14) generator\ \cite{Mangano:2002ea} coupled with \Pythia (v6.42)
and CTEQ6L1\ PDF sets\ \cite{Pumplin:2002vw}. The latter is also used
to test the $\Wjets$ modelling. The diboson processes ($\WW$, $\WZ$,
$\ZZ$) were simulated using the \Herwig\ (v6.52) and \Jimmy\ (v4.31)
generators with the AUET2 tune\ \cite{ATL-PHYS-PUB-2011-008} and the
CTEQ6L1\ PDF set. The single-boson and diboson samples were
normalized to their production cross-sections calculated at
next-to-next-to-leading order\ (NNLO)\
\cite{Anastasiou:2003ds} and NLO\ \cite{Campbell:2011bn},
respectively. The multi-jet background was modelled by a data-driven
method as described in Section~\ref{sec:bkg}.

Almost all the generated event samples were passed through the full
\ATLAS detector simulation\ \cite{ATL-2010-005} based on \Geant{}4\
\cite{Agostinelli:2002hh} and then processed with the same
reconstruction chain as the data. The remaining samples, which consist
of single-top-quark $s$- and $t$-channel samples for scale variation
studies, as well as for $t$-channel and $\ttbar$ modelling studies,
are passed through the ATLFAST2 simulation of the \ATLAS detector,
which uses a fast simulation for the calorimeters\
\cite{ATLAS:2010bfa}. The simulated events were overlaid with
additional minimum-bias events generated with \Pythia to simulate the
effect of additional $\pp$ interactions. All processes involving top
quarks were generated using a top-quark mass of $\unit[172.5]{\GeV}$.

%-------------------------------------------------------------------------------

\section{Background estimation}
\label{sec:bkg}

The two most important backgrounds are $\ttbar$ and $\Wjets$
production. The former is difficult to distinguish from the signal
since $\ttbar$ events contain real top-quark decays. In its dileptonic
decay mode, $\ttbar$ events can mimic the final-state signature of the
signal if one of the two leptons escapes unidentified, whereas the
semileptonic decay mode contributes to the selected samples if only
two of its four jets are identified or if some jets are merged. The
$\Wjets$ events can contribute to the background if they contain
$\bquark$-jets in the final state or due to mis-tagging of jets
containing other quark flavours. Single top-quark $t$-channel
production also leads to a sizeable background contribution, while
associated $\Wt$ production has only a small effect.

A less significant background contribution is multi-jet production
where jets, non-prompt leptons from heavy-flavour decays, or electrons
from photon conversions are mis-identified as prompt isolated leptons.
This background is estimated by using a data-driven matrix method\
\cite{TopFakes}, where the probability to mis-identify an isolated
electron or muon in an event is obtained by exploiting sum rules based
on disjoint control samples, one almost pure electron or muon sample
and another containing a high fraction of mis-identified leptons due
to a relaxed lepton-isolation criterion. For both decay channels the
amount of multi-jet background is below 2\% in the final selection.
Other minor backgrounds are from $\Zjets$ and diboson production.

Apart from the data-driven multi-jet background, all samples are
normalized to their predicted cross-sections. The samples for single
top-quark production are normalized to their NLO+NNLL predictions (see
Section\ \ref{sec:intro}), while for all $\ttbar$ samples a recent
calculation with Top++\ (v2.0) at NNLO in QCD including resummations
of NNLL soft gluon terms of
$\sigma^{\text{th}}_{\ttbar}\!=\!\unit[253^{+13}_{-15}]{pb}$ is used
for the normalization \
\cite{Cacciari:2011hy,Baernreuther:2012ws,Czakon:2012zr,%
  Czakon:2012pz,Czakon:2013goa,Czakon:2011xx}.

%-------------------------------------------------------------------------------

\section{Event reconstruction and selection}
\label{sec:selection}

For the selection of $s$-channel final states, a single high-$\Pt$
lepton, either electron or muon, exactly two $b$-tagged jets and a
large amount of $\Etmiss$ are required.

Electrons are reconstructed as energy deposits in the electromagnetic
calorimeter matched to charged-particle tracks in the inner detector
and must pass tight identification requirements\
\cite{Aad:2011mk,Aad:2014fxa}. The transverse momentum of the
electrons must satisfy $\Pt\!>\!\unit[30]{\GeV}$ and be in the central
region with pseudorapidity $\abseta\!<\!2.47$, excluding the region
$1.37\!<\!\abseta\!<\!1.52$, which contains a large amount of inactive
material. Muon candidates are identified using combined information
from the inner detector and the muon spectrometer\
\cite{Aad:2014zya,Aad:2014rra}. They are required to have
$\Pt\!>\!\unit[30]{\GeV}$ and $\abseta\!<\!2.5$. Both the electrons
and muons must fulfil additional isolation requirements, as described
in Ref.~\cite{TopFakes}, in order to reduce contributions from
non-prompt leptons originating from hadron decays, and fake leptons.

Jets are reconstructed by using the anti-$k_t$ algorithm\
\cite{Cacciari:2008gp} with a radius parameter of $0.4$ for
calorimeter energy clusters calibrated with the local cluster
weighting method\ \cite{Aad:2011he}. For the jet calibration an
energy- and $\eta$-dependent simulation-based scheme with in-situ
corrections based on data\ \cite{Aad:2014bia} is employed. Only events
containing exactly two jets with $\Pt\!>\!\unit[40]{\GeV}$ for the
leading jet and $\Pt\!>\!\unit[30]{\GeV}$ for the second leading jet,
as well as $\abseta\!<\!2.5$ for both jets are selected. Events
involving additional jets with $\Pt\!>\!\unit[25]{\GeV}$ and
$\abseta\!<\!4.5$ are rejected. Both jets must be identified as
$\bquark$-jets. The identification is performed using a neural network
which combines spatial and lifetime information from secondary
vertices of tracks associated with the jets. The operating point of
the tagging algorithm used in this analysis corresponds to a
$\bquark$-tagging efficiency of 70\% and a rejection factor for
light-flavour jets of about 140, while the rejection factor for charm
jets is around\ 5\ \cite{btag1,btag2}.

The missing transverse momentum is computed from the vector sum of all
clusters of energy deposits in the calorimeter that are associated
with reconstructed objects, and the transverse momenta of the
reconstructed muons\ \cite{Aad:2012re,ATLAS-CONF-2013-082}. The energy
deposits are calibrated at the corresponding energy scale of the
parent object. Since $\Etmiss$ is a measure for the undetectable
neutrino originating from the top-quark decay, in this analysis only
events with $\Etmiss\!>\!\unit[35]{\GeV}$ are accepted. Furthermore,
the transverse mass\footnote{The transverse mass, $\mtW$, is computed
  from the lepton transverse momentum, $\PtLep$, and the difference in
  azimuthal angle, $\varDelta\phi$, between the lepton and the
  missing transverse momentum as
  $\mtW=\sqrt{2\Etmiss\PtLep\left(1-\cos(\varDelta\phi(\hat
      E_{\text{T}}^{\text{miss}}, \,\hat
      p_{\text{T}}^{\ell}))\right)}\,$.} of the $\Wboson$ boson,
$\mtW$, needs to be larger than $\unit[30]{\GeV}$ to suppress
multi-jet background.

The main background at this stage of the selection originates from
top-quark pair production, which is in turn dominated by dileptonic
$\ttbar$ events. To reduce this background, a veto is applied to all
events containing an additional reconstructed electron or muon
identified with loose criteria. The minimum required $\Pt$ of these
leptons is $\unit[5]{\GeV}$. By this measure the $\ttbar$ background
is diminished by 30\% while reducing the signal by less than half a
per cent. After the application of all event selection criteria, a
signal-to-background ratio of 4.6\% is reached. The event yields for
all samples in the signal region are collected in
Table~\ref{tab:postfit-yields-sig}.

Apart from the signal region, two more regions are defined to validate
the modelling, one validation region for $\ttbar$ production and a
control region for the $\Wjets$ background. The latter is used to
constrain the normalization of the $\Wjets$ background in the final
signal extraction, as explained in more detail in
Section~\ref{sec:extraction}. The two regions are defined in the same
way as the signal region, except that neither the veto on events with
additional jets nor the one on dilepton events are applied. Top-quark
pair production is enriched by selecting events containing exactly
four jets with $\Pt\!>\!\unit[25]{\GeV}$, two of them $\bquark$-tagged
at the 70\% working point. The $\Wjets$ control region is defined
using a less stringent $\bquark$-tag requirement (80\% working point);
in order to ensure that this region is disjoint from the signal
region, it is required that at least one of the two jets fails to meet
the signal region $\bquark$-tagging criteria at the 70\% working
point.

%-------------------------------------------------------------------------------

\section{Matrix element method}
\label{sec:mem}

The ME method directly uses theoretical calculations to compute a
per-event signal probability. This technique was used for the
observation of single top-quark production at the \Tevatron\
\cite{Abazov:2006gd,Aaltonen:2009jj,Abazov:2009ii}. The discrimination
between signal and background is based on the computation of
likelihood values $\mathcal{P}(X|H_{\text{proc}})$ for the hypothesis
that a measured event with final state $X$ is of a certain process
type $H_{\text{proc}}$. Those likelihoods can be computed by means of
the factorization theorem from the corresponding partonic
cross-sections of the hard scatter. The mapping between the hadronic
measured final state and the parton state is implemented by transfer
functions which take into account the detector resolution functions, the
reconstruction and $\bquark$-tagging efficiencies, as well as all
possible permutations between the partons and the reconstructed
objects.

The phase-space integration of the differential partonic
cross-sections is performed using the Monte Carlo integration
algorithm \Vegas\ \cite{Lepage:1977sw} from the \Cuba program library\
\cite{Hahn:2005pf}. The required PDF sets are taken from the
\LHAPDF{}5 package\ \cite{Whalley:2005nh}, while the computation of
the scattering amplitudes is based on codes from the \MCFM program\
\cite{mcfm}. The parameterizations of the \ATLAS detector resolutions
used for the transfer functions are those used in the KLFitter
kinematic fit framework\ \cite{Erdmann:2013rxa,ATLAS:2012an}.

In total, eight different processes are considered for the computation
of the likelihood values. For the $s$-channel signal, final states
with two and three partons are included, while the single-top-quark
$t$-channel process is modelled in the four-flavour scheme only. In
the case of $\ttbar$ production semileptonic and dileptonic processes
are evaluated separately. The remaining background processes are
$\Wboson$ boson production with two associated light-flavour jets,
with one light-flavour jet and one charm jet, and $\Wplusbb$
production.

From the likelihood values of these processes the probability $P(S|X)$
for a measured event $X$ to be a signal event $S$ can be computed with
Bayes' theorem by
\begin{equation}
  \label{eq:me_discr}
  P(S|X) = \frac{\sum_i\alpha_{S_i}\mathcal{P}(X|S_i)}
  {\sum_i\alpha_{S_i}\mathcal{P}(X|S_i)
    + \sum_j\alpha_{B_j}\mathcal{P}(X|B_j)}\,.
\end{equation}
Here, $S_i$ and $B_j$ denote all signal and background processes that
are being considered. The \textit{a priori} probabilities
$\alpha_{S_i}$ and $\alpha_{B_j}$ are given by the expected fraction
of events of each process in the set of selected events within the
signal region. The value of $P(S|X)$ is taken as the main discriminant
in the signal extraction. The binning of the ME discriminant is
optimized in the signal region utilizing a dedicated algorithm\
\cite{Aad:2014xzb}. This results in a non-equidistant binning which
exhibits wider bins in regions with a large signal contribution, while
preserving a sufficiently large number of background events in each
bin. In the following, all histograms showing the ME\ discriminant are
drawn with a constant bin width causing a non-linear horizontal scale.
Values of $P(S|X)$ lower than $0.00015$ are not taken into account for
the signal extraction because of the large background domination in
this range.

\begin{figure*}
  \centering
  \subfloat[$\Wjets$ region]{%
    \includegraphics[width=0.46\textwidth]{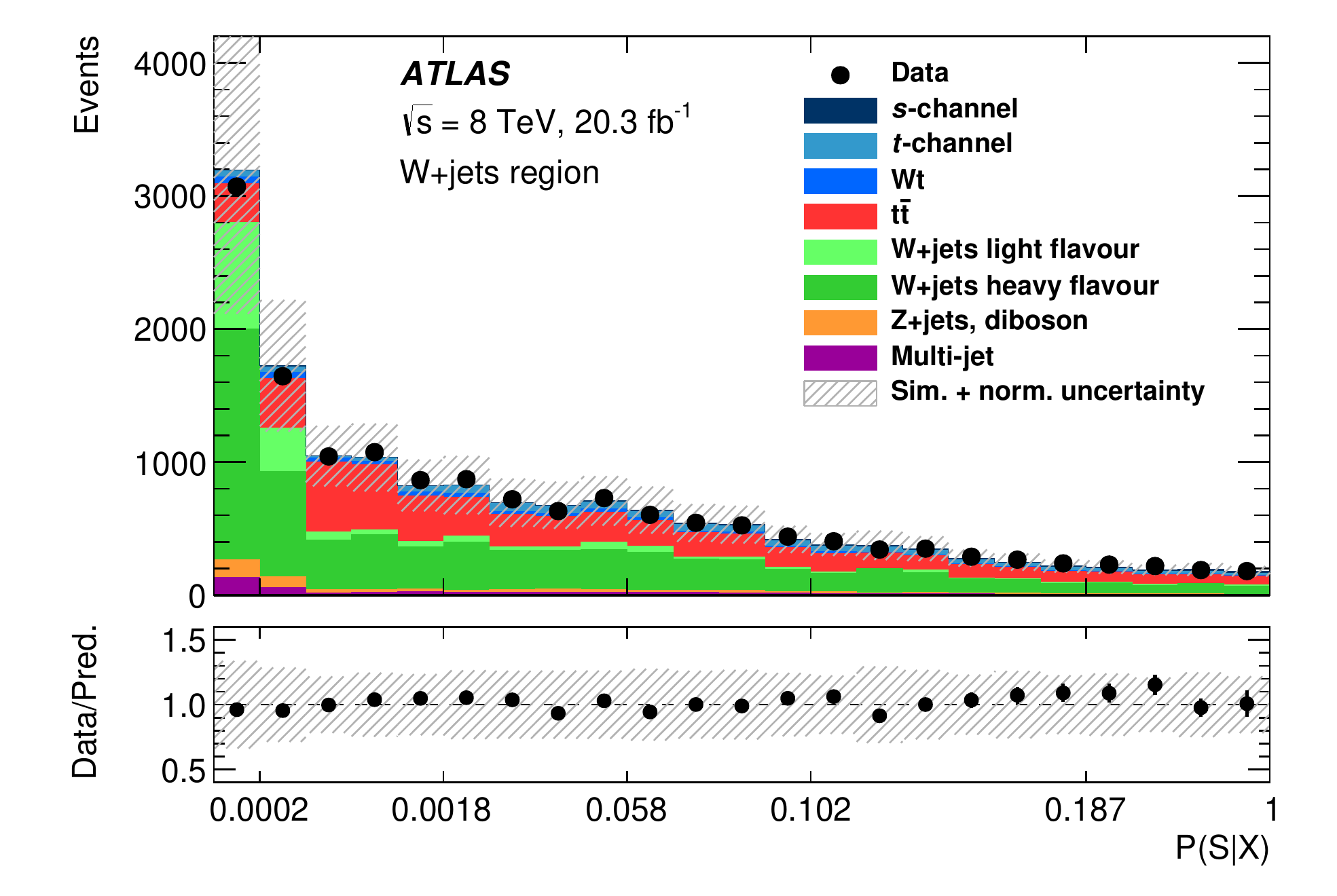}
    \label{fig:mediscr_ctrl-wjets}
  } \qquad
  \subfloat[$\ttbar$ region]{%
    \includegraphics[width=0.46\textwidth]{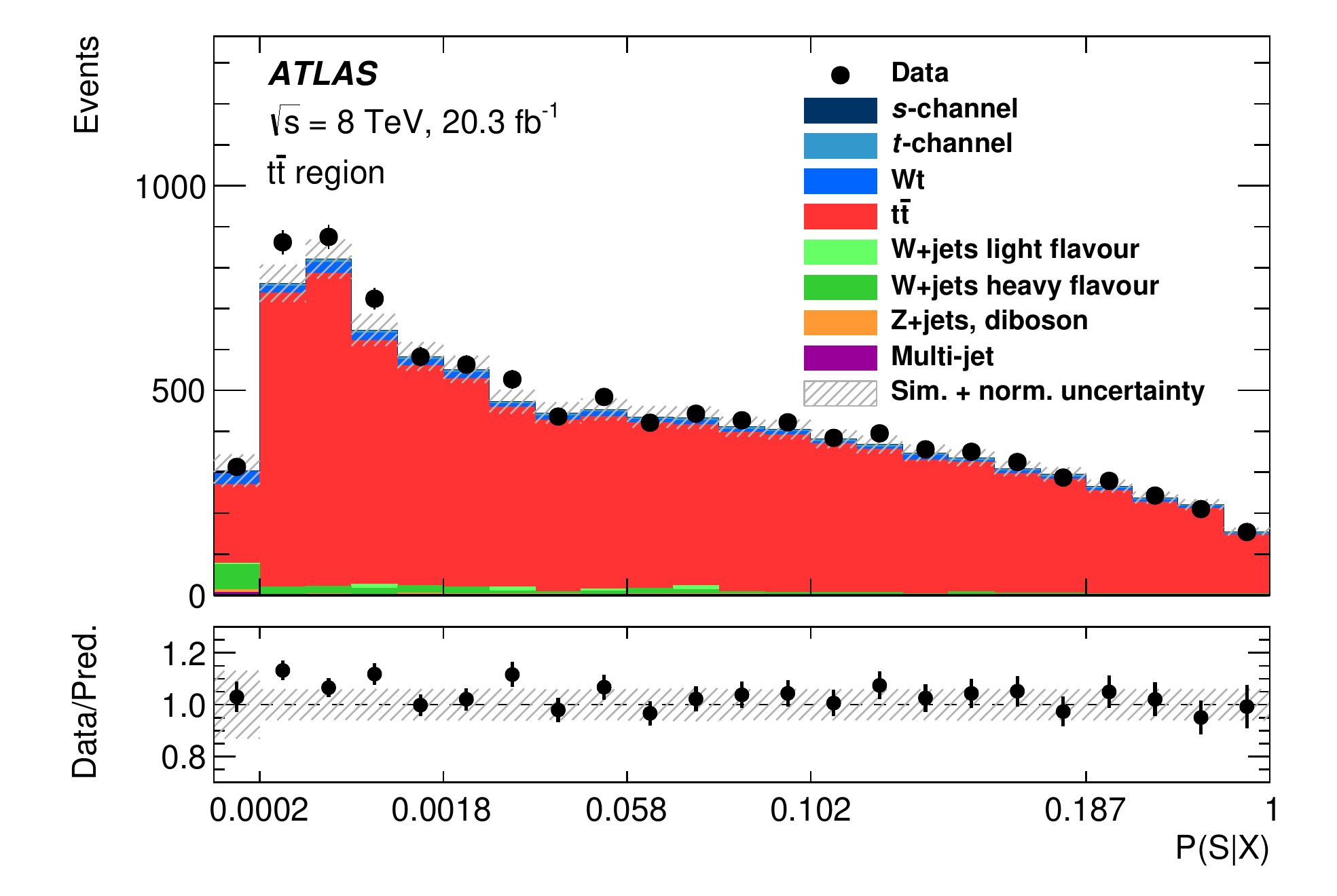}
    \label{fig:mediscr_ctrl-ttbar}
  }
  \caption{%
    ME discriminant for the \protect\subref{fig:mediscr_ctrl-wjets}
    $\Wjets$ control region and
    \protect\subref{fig:mediscr_ctrl-ttbar} for the $\ttbar$
    validation region. Except for the data-driven multi-jet background
    all samples are scaled to their SM expectation. The hatched bands
    represent the predicted normalization uncertainty of all processes
    added in quadrature with the statistical uncertainty of the
    simulation. The lower panels show the relative difference between
    data and the prediction. Both histograms use the non-equidistant
    binning of the ME discriminant $P(S|X)$ optimized for the signal
    region.}
  \label{fig:mediscr_ctrl}
\end{figure*}

In order to validate the $s$-channel ME discriminant $P(S|X)$ a
comparison of the discriminant between data and the simulation is
shown in Fig.~\ref{fig:mediscr_ctrl} for the $\Wjets$ control region
and the $\ttbar$ validation region. For the latter, only the two
$\bquark$-tagged jets are considered for the ME discriminant
computation. The normalization of each sample in
Fig..~\ref{fig:mediscr_ctrl} except for the data-driven mulit-jet
background is obtained by a similar fit to data as described in
Sec.~\ref{sec:extraction}. The only difference is that here all
samples, including the signal, are varied within their SM prediction
unctertainy, while for the signal extraction fit the signal is allowed
to float freely. In both regions the data is described well by the
simulation.

%-------------------------------------------------------------------------------

\section{Systematic uncertainties}
\label{sec:systematics}

Apart from systematic effects in the signal acceptance and the
background normalizations, the ME discriminant is subject to those
systematic effects which change the four-momenta of the reconstructed
objects. Therefore, systematic uncertainties such as the energy
calibration of jets, electrons and muons are propagated through the
whole analysis including the ME computation by variations in the
modelling of the detector response.

The main sources of systematic uncertainties for jets are the energy
scale, which is evaluated by a combination of in-situ techniques\
\cite{Aad:2014bia}, the energy resolution\ \cite{Aad:2012ag} and the
reconstruction efficiency\ \cite{Aad:2014bia}. For $\bquark$,
$\cquark$ and light-flavour tagging of the jets the modelling of the
respective efficiencies is taken into account \ \cite{btag1,btag2}.
The lepton uncertainties originate from trigger, identification and
isolation efficiencies, as well as from their energy scale and
resolution\ \cite{Aad:2014fxa,Aad:2014rra}. Both the energy scale and
energy resolution uncertainties for the jets and leptons are
transferred to the $\Etmiss$ calculation. The impact of low-$\Pt$ jets
on $\Etmiss$ and contributions from energy deposits in the calorimeters
not associated with any reconstructed objects are considered as well.

Potential mis-modelling in the simulation of the signal and the main
background processes is also taken into account in the evaluation of
the systematic uncertainties. This includes contributions from the
modelling of the hard process, the parton showers, hadronization and
ISR/FSR. The uncertainty caused by the choice of renormalization and
factorization scales is evaluated for the signal process and $\ttbar$
production. All of these uncertainties are estimated by comparing
simulation samples produced with different generators (see
Section~\ref{sec:samples}) or different parameter settings such as
shower models or scales.

The normalization uncertainties of the different samples are taken
from theory except for the multi-jet background, which is estimated by
a data-driven method. Uncertainties of 6\%, 4\% and 7\% are assigned to
$\ttbar$, single top-quark $t$-channel, and $\Wt$ production,
respectively. For $\Wjets$ production, as well as for the combination
of $\Zjets$ and diboson production, an uncertainty of 60\% each is
considered\ \cite{Ellis:1985vn,Berends:1990ax}. The uncertainty for
$\Wjets$ production is dominated by its heavy-flavour contribution.
For the multi-jet background a normalization uncertainty of 50\% is
estimated.

The uncertainties associated with the PDFs are taken into account for
all simulated samples by assessing a systematic uncertainty according
to the PDF4LHC prescription\ \cite{Botje:2011sn}, which makes use of
the MSTW2008NLO\ \cite{Martin:2009iq}, CT10, and NNPDF2.3\
\cite{Ball:2012cx} sets. The uncertainty of the luminosity measurement
is 2.8\%, which was determined by dedicated beam-separation scans\
\cite{Aad:2013ucp}.

In addition to the impact of the systematic uncertainties on the
signal acceptance and the background normalizations, their effect on
the shape of the discriminant distributions is taken into account if
it is significant. The significance is evaluated by performing
$\chi^2$ tests between the nominal and the systematically varied
distributions made from uncorrelated event samples. Only a small
fraction of all systematic uncertainties exhibit a significant shape
effect. These are mainly the impact of the jet energy scale and
resolution on the single-top-quark $s$- and $t$-channel samples.

For all simulation samples the effect of their limited size is
included in the systematic uncertainty.

%-------------------------------------------------------------------------------

\section{Signal extraction}
\label{sec:extraction}

The amount of signal in the selected data set is measured by means of
a binned maximum-likelihood fit of the ME discriminant in the signal
region. In order to better constrain the $\Wjets$
background, the lepton charge in the $\Wjets$-enriched control region
is used as an additional discriminant variable in the fit, as it exploits
the charge asymmetry of the incoming partons participating in the
$\Wjets$ processes. The likelihood function used in the fit consists
of a Poisson term for the overall number of observed events, a product
of probability densities of the discriminants taken over all bins of
the distributions and a product of Gaussian constraint terms for the
nuisance parameters which incorporate all statistical and systematic
uncertainties in the fit. While all backgrounds are constrained by
their given uncertainties, the signal strength
$\mu\!=\!\sigma_s/\sigma^{\text{th}}_s$ is a free parameter in the
fit.

The significance of the fit result is obtained with a
profile-likelihood-ratio test statistic which is used to determine how
well the fit result agrees with the background-only hypothesis.
Ensemble tests for all nuisance parameters are performed using the
aforementioned likelihood function to get the expected distributions
of the test statistic for the background-only and the
signal-plus-background hypotheses. The significance is evaluated by
integrating the probability density of the test statistic expected for
the background-only hypothesis above the observed value. In a similar
fashion the confidence interval of the measured signal strength can be
estimated by studying its $p$-value dependence for the background-only
hypothesis, as well as for the signal-plus-background hypothesis, by
means of ensemble tests. The statistical evaluation used throughout
this analysis is based on the RooStats framework\
\cite{Moneta:2010pm}.

%-------------------------------------------------------------------------------

\section{Results}
\label{sec:result}

\begin{figure*}
  \centering
  \subfloat[ME discriminant]{%
    \includegraphics[width=0.46\textwidth]{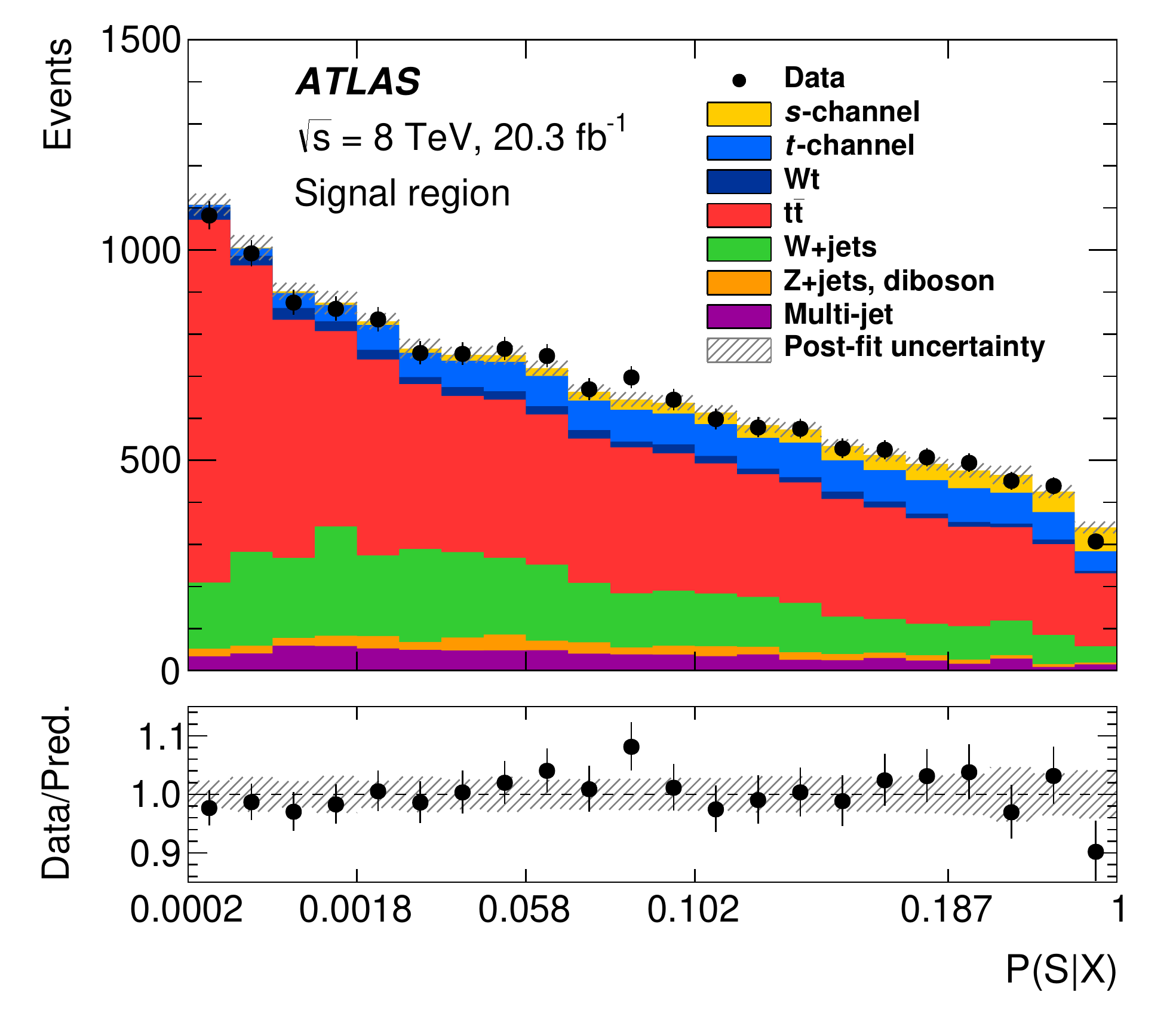}
    \label{fig:postfit-me-obs}
  } \qquad
  \subfloat[Lepton charge discriminant]{%
    \includegraphics[width=0.46\textwidth,page=1]{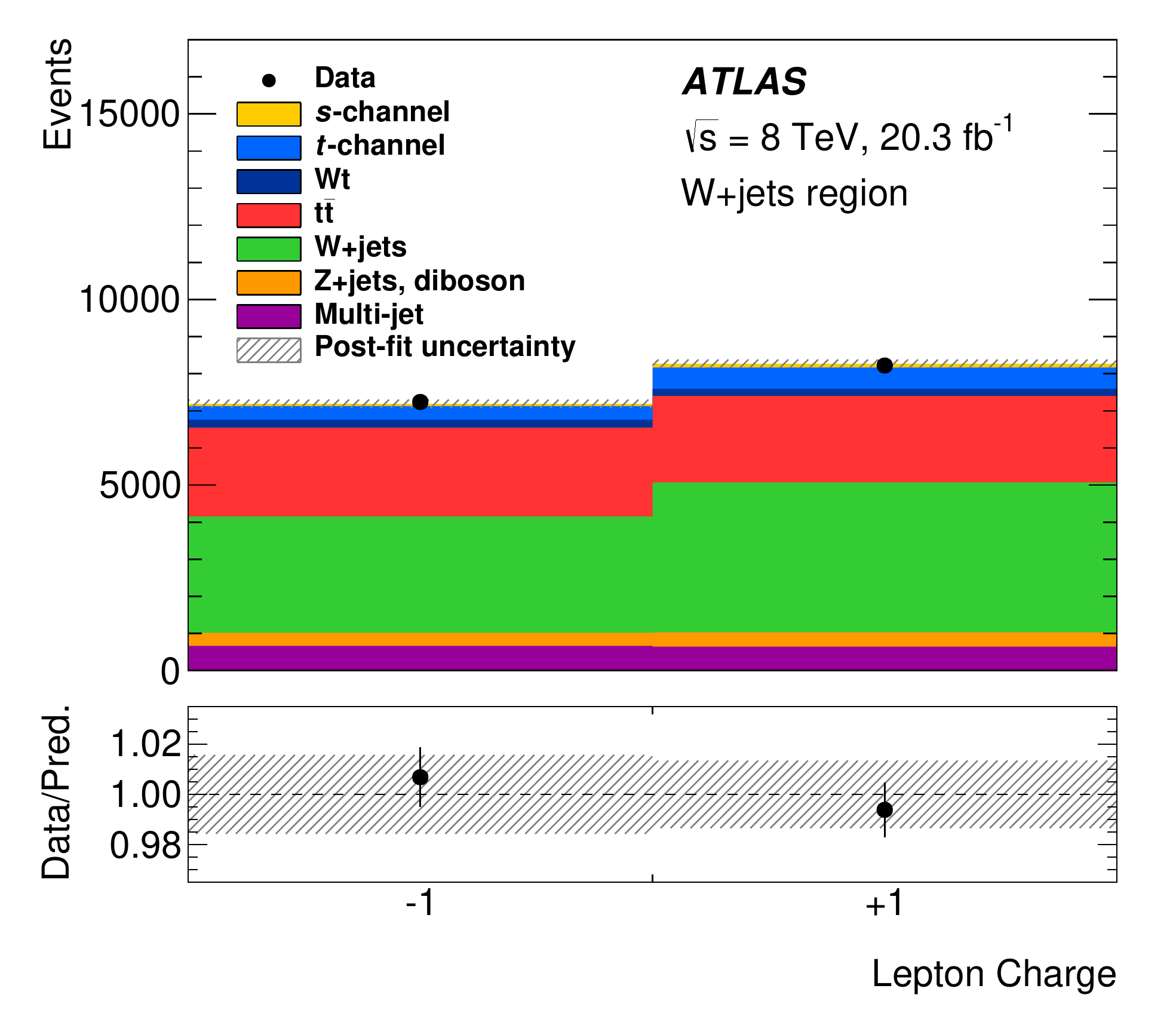}
    \label{fig:postfit-lepchg-obs}
  } 
  \caption{Post-fit distribution of
    \protect\subref{fig:postfit-me-obs} the ME discriminant in the
    signal region and \protect\subref{fig:postfit-lepchg-obs} the
    lepton charge discriminant in the $\Wjets$ control region. All
    samples are scaled by the fit result utilizing all fit parameters.
    The hatched bands indicate the total uncertainty of the post-fit
    result including all correlations. The ME distributions are made
    using the optimized, non-equidistant binning which is also applied
    in the signal extraction fit.}
  \label{fig:postfit-obs}
\end{figure*}

\begin{figure}
 \centering
 \includegraphics[width=0.48\textwidth]{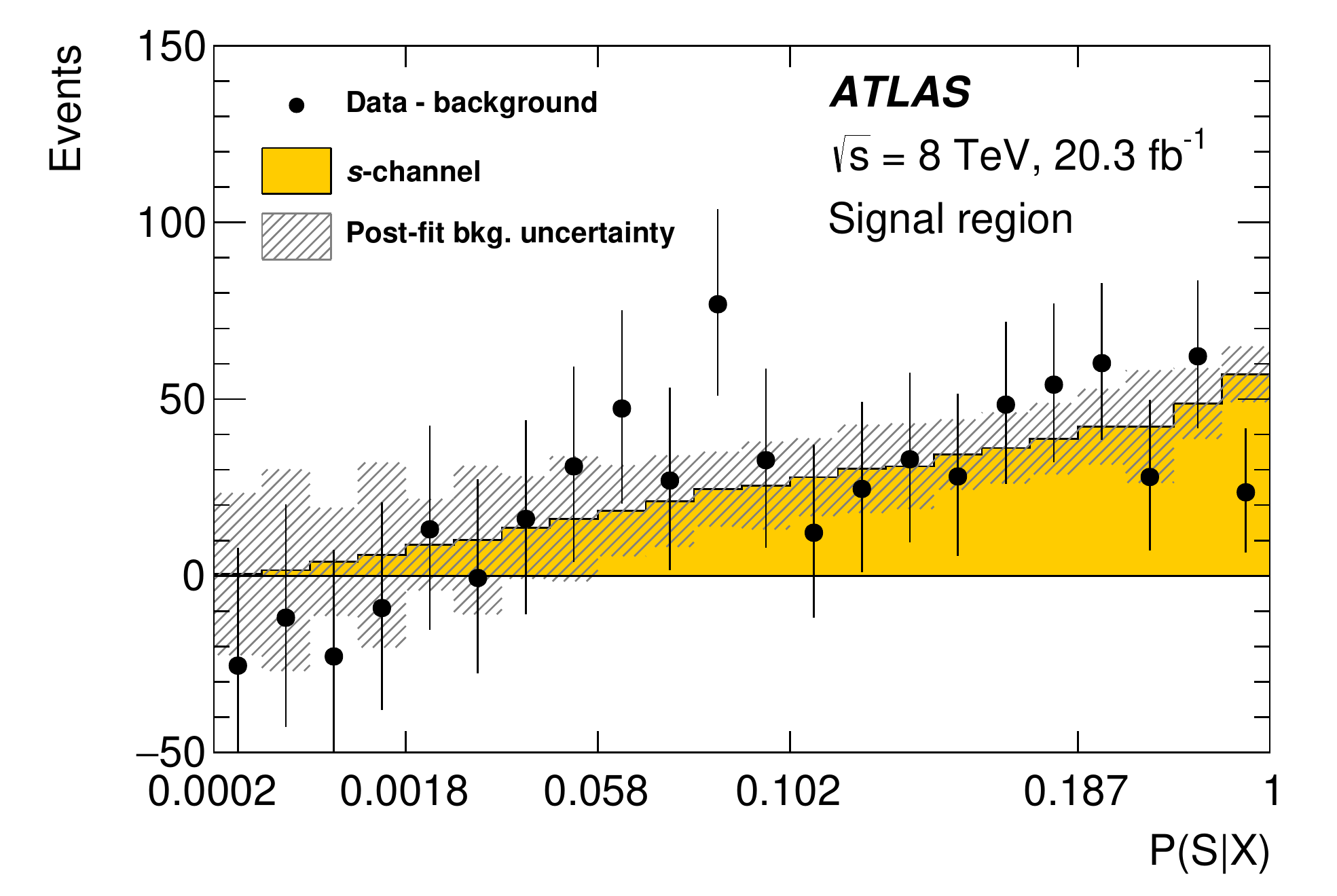}
 \caption{Distribution of the ME discriminant in data in the signal
   region after the subtraction of all backgrounds (post-fit), showing
   the signal contribution. The error bars indicate the uncertainty of
   the measurement in each bin. The fitted distribution for the
   simulation of the signal is also shown together with its fit
   uncertainty for all backgrounds given by the hatched band. The
   binning is the same as the optimized, non-equidistant binning used
   in the fit.}
  \label{fig:postfit-me-diff-obs}
\end{figure}

The results of the maximum-likelihood fit are presented in
Fig.~\ref{fig:postfit-obs}, which shows the two discriminant
distributions used in the fit for all samples scaled by the fit
results. For the ME discriminant the signal contribution in the data
after the subtraction of all background samples is given in
Fig.~\ref{fig:postfit-me-diff-obs}. After the fit, none of the nuisance
parameters is biased or further constrained by the fit, except for the
$\Wjets$ normalization. Here, the rather conservative input
uncertainty is halved by the fit to signal and the $\Wjets$ control
regions. The observed signal strength obtained by the fit is
$\mu\!=\!0.86^{+0.31}_{-0.28}$ with an observed (expected)
significance of 3.2\ (3.9) standard deviations.
Table~\ref{tab:postfit-yields-sig} summarizes the pre-fit and post-fit
event yields for the signal and all backgrounds.


\begin{table}
  \centering
  \renewcommand{\arraystretch}{1.1}
  \sisetup{round-mode=figures, round-precision=2,
    retain-explicit-plus=true}
  \begin{tabular}{%
      l@{\qquad}
      S[table-number-alignment = right,
      table-figures-decimal = 0]
      S[table-number-alignment = right,
      table-figures-decimal = 0]@{\,}@{$\pm$}@{\,}
      S[table-number-alignment = left,
      table-figures-integer = 3]@{\quad}
      S[table-number-alignment = center,
      table-figures-decimal = 3,
      round-precision = 3]
    }
    \toprule
    Process & \multicolumn{1}{c}{Pre-fit} & \multicolumn{2}{c}{Post-fit} & \multicolumn{1}{c}{Post-fit\,/\,Pre-fit} \\
    \midrule 
    Single-top $s$-channel & {\phO\numRF{605.955}{2}} & {\phO\phdOO\numRF{538.75}{2}} & {\numRF{161.532}{2}} & {\phO\numRF{0.89}{2}} \\
    Single-top $t$-channel & {\numRF{1233.0 }{3}} & {\phdOO\numRF{1362.85}{3}} & {\numRF{156.42}{2}} & 1.103 \\
    Assoc. $\Wt$ production & {\phO\numRF{370.523}{2}} & {\phO\phdOO\numRF{383.449}{2}} & {\phO\numRF{51.0005}{1}} & 1.024 \\
    $\ttbar$ production & {\numRF{8213.68}{2}} & {\phdOO\numRF{8084.11}{2}} & {\numRF{376.45}{1}} & {\phO\numRF{0.986}{2}} \\
    $\Wjets$ & {\numRF{2645.68}{2}} & {\phdOO\numRF{3085.95}{2}} & {\numRF{512.944}{1}} & 1.1698 \\
    $\Zjets$ \& diboson & {\phO\numRF{292.933}{2}} & {\phO\phdOO\numRF{411.688}{2}} & {\numRF{277.199}{2}} & 1.3993 \\
    Multi-jet & {\phO\numRF{622.823}{1}} & {\phO\phdOO\numRF{800.451}{1}} & {\numRF{364.945}{1}} & 1.284 \\
    \midrule
    Total expectation & {\numRF{13984.7}{4}\phdO} & {\phO\numRF{14667.3}{4}} & {\numRF{180.968}{2}} & 1.0493 \\
    Data & \multicolumn{2}{c}{\numRF{14677}{5}\phO} &  &  \\
    \bottomrule
  \end{tabular}
  \caption{Pre-fit and post-fit event yields in the signal region for
    ME discriminant values larger than $0.00015$. The
    post-fit uncertainty corresponds to the uncertainty of the
    maximum-likelihood fit including all correlations. The last column shows
    the ratio of post-fit to pre-fit results.}
  \label{tab:postfit-yields-sig}  
\end{table}

This analysis measures a cross-section of
$\sigma_{s}\!=\!\unit[4.8\pm0.8\text{(stat.)}
^{+1.6}_{-1.3}\text{(syst.)}]{pb}\!=\!\unit[4.8^{+1.8}_{-1.6}]{pb}$.
The main sources of uncertainty are collected in
Table~\ref{tab:syst-breakdown}. The largest contribution arises from
the limited sample sizes for data and the simulation. The jet energy
resolution plays a major role, as well as the modelling of the
single-top-quark $t$-channel background and scale variations for the
signal. All other systematic effects are negligible.
\begin{table}
  \centering
  \begin{tabular}{l%
      S[table-number-alignment = right,
        input-comparators,
        round-precision = 0,
        round-mode=places,
        table-figures-decimal = 0]}
    \toprule
    Type & \multicolumn{1}{c}{$\pm\varDelta\sigma/\sigma$ [\%]} \\
    \midrule
    Data statistics & 15.755 \\
    \midrule
    MC statistics & 12.221 \\
    Jet energy resolution & 11.988 \\
    $t$-channel generator choice & 10.913 \\
    $\bquark$-tagging & 7.584 \\
    $s$-channel generator scale & 7.283 \\
    $\Wjets$ normalization & 6.088 \\
    Luminosity & 5.262 \\
    $t$-channel normalization & 5.061 \\
    Jet energy scale & 4.788 \\
    PDF & 2.524 \\
    Lepton identification & 2.345 \\
    Electron energy scale & 1.453 \\
    $\ttbar$ generator choice & 1.409 \\
    Lepton trigger & 1.401 \\
    Charm tagging & 1.118 \\
    Other & <1.0 \\
    \midrule
    Total & 33.652 \\
    \bottomrule
  \end{tabular}
  \caption{Main statistical and systematic uncertainties contributing
    to the total uncertainty of the measured cross-section. The
    relative uncertainties reflect the influence of each systematic
    effect on the overall signal strength uncertainty. Apart from
    possible correlations between the systematic uncertainties, the
    total uncertainty contains several minor contributions which are
    all smaller than 1\%.}
  \label{tab:syst-breakdown}
\end{table}

The measured cross-section can be interpreted in terms of the CKM
matrix element $\Vtb$. The ratio of the measured cross-section to the
prediction is equal to $|\fLV\Vtb|^2$, where the form factor $\fLV$
could be modified by new physics or radiative corrections through
anomalous coupling contributions, for example those in
Refs.~\cite{AguilarSaavedra:2008zc,Kane:1991bg,Rizzo:1995uv}. The
$s$-channel production and top quark decays through $\absVts$ and
$\absVtd$ are assumed to be small. A lower limit on $\absVtb$ is
obtained for $\fLV\!=\!1$ as in the SM, without assuming CKM
unitarity\ \cite{Alwall:2006bx,Cao:2015qta}. The measured value of
$|\fLV\Vtb|$ is $0.93^{+0.18}_{-0.20}$, and the corresponding lower
limit on $\absVtb$ at the 95\% confidence level is 0.5.

%-------------------------------------------------------------------------------
\FloatBarrier
\section{Conclusion}
\label{sec:conclusion}

An analysis for $s$-channel single top-quark production in $\pp$
collisions at a centre-of-mass energy of $\unit[8]{\TeV}$ recorded by
the \ATLAS detector at the \LHC is presented. The analysed data set
corresponds to an integrated luminosity of $\unit[20.3]{\ifb}$. The
selected events consist either of an electron or muon, two jets, both
of which are identified to be induced by a $\bquark$-quark, and
large $\Etmiss$. In order to separate the signal from the large
background contributions, a matrix element method discriminant is
used. The signal is extracted from the data utilizing a profile
likelihood fit, which leads to a measured cross-section of
$\unit[4.8^{+1.8}_{-1.6}]{pb}$. The result, which is in agreement with
the SM prediction, corresponds to an observed significance of 3.2\
standard deviations, while the expected significance of the analysis
is 3.9\ standard deviations.

%-------------------------------------------------------------------------------
\section*{Acknowledgements}
%-------------------------------------------------------------------------------

% Acknowledgements for papers with collision data
% Version 15-Sep-2015

% Standard acknowledgements start here
%----------------------------------------------
We thank CERN for the very successful operation of the LHC, as well as the
support staff from our institutions without whom ATLAS could not be
operated efficiently.

We acknowledge the support of ANPCyT, Argentina; YerPhI, Armenia; ARC, Australia; BMWFW and FWF, Austria; ANAS, Azerbaijan; SSTC, Belarus; CNPq and FAPESP, Brazil; NSERC, NRC and CFI, Canada; CERN; CONICYT, Chile; CAS, MOST and NSFC, China; COLCIENCIAS, Colombia; MSMT CR, MPO CR and VSC CR, Czech Republic; DNRF, DNSRC and Lundbeck Foundation, Denmark; IN2P3-CNRS, CEA-DSM/IRFU, France; GNSF, Georgia; BMBF, HGF, and MPG, Germany; GSRT, Greece; RGC, Hong Kong SAR, China; ISF, I-CORE and Benoziyo Center, Israel; INFN, Italy; MEXT and JSPS, Japan; CNRST, Morocco; FOM and NWO, Netherlands; RCN, Norway; MNiSW and NCN, Poland; FCT, Portugal; MNE/IFA, Romania; MES of Russia and NRC KI, Russian Federation; JINR; MESTD, Serbia; MSSR, Slovakia; ARRS and MIZ\v{S}, Slovenia; DST/NRF, South Africa; MINECO, Spain; SRC and Wallenberg Foundation, Sweden; SERI, SNSF and Cantons of Bern and Geneva, Switzerland; MOST, Taiwan; TAEK, Turkey; STFC, United Kingdom; DOE and NSF, United States of America. In addition, individual groups and members have received support from BCKDF, the Canada Council, CANARIE, CRC, Compute Canada, FQRNT, and the Ontario Innovation Trust, Canada; EPLANET, ERC, FP7, Horizon 2020 and Marie Skłodowska-Curie Actions, European Union; Investissements d'Avenir Labex and Idex, ANR, Region Auvergne and Fondation Partager le Savoir, France; DFG and AvH Foundation, Germany; Herakleitos, Thales and Aristeia programmes co-financed by EU-ESF and the Greek NSRF; BSF, GIF and Minerva, Israel; BRF, Norway; the Royal Society and Leverhulme Trust, United Kingdom.

The crucial computing support from all WLCG partners is acknowledged
gratefully, in particular from CERN and the ATLAS Tier-1 facilities at
TRIUMF (Canada), NDGF (Denmark, Norway, Sweden), CC-IN2P3 (France),
KIT/GridKA (Germany), INFN-CNAF (Italy), NL-T1 (Netherlands), PIC (Spain),
ASGC (Taiwan), RAL (UK) and BNL (USA) and in the Tier-2 facilities
worldwide.
%----------------------------------------------

%The \texttt{atlaslatex} package contains the acknowledgements that were valid 
% at the time of the release you are using.
% These can be found in the \texttt{acknowledgements} subdirectory.
% When your ATLAS paper or PUB/CONF note is ready to be published,
% download the latest set of acknowledgements from:\\
% \url{https://twiki.cern.ch/twiki/bin/view/AtlasProtected/PubComAcknowledgements}

% The supporting notes for the analysis should also contain a list of contributors.
% This information should usually be included in \texttt{mydocument-metadata.tex}.
% The list should be printed either here or before the table of contents.

%-------------------------------------------------------------------------------
%\clearpage
%\appendix
%\part*{Appendix}
%\addcontentsline{toc}{part}{Appendix}
%-------------------------------------------------------------------------------
%
%In a paper, an appendix is used for technical details that would otherwise disturb the flow of the paper.
%Such an appendix should be printed before the Bibliography.

%-------------------------------------------------------------------------------
% If you use biblatex and either biber or bibtex to process the bibliography
% just say \printbibliography here
\printbibliography
% If you want to use the traditional BibTeX you need to use the syntax below.
%\bibliographystyle{bibtex/bst/atlasBibStyleWoTitle}
%\bibliography{TOPQ_2015_XX,bibtex/bib/ATLAS}
%-------------------------------------------------------------------------------

\newpage
% ATLAS Collaboration author list
% Data extracted on 21-Oct-2015 for paper reference TOPQ-2015-01
%\documentclass[11pt]{article}
%\usepackage{a4wide}\begin{document}
\begin{flushleft}
{\Large The ATLAS Collaboration}

\bigskip

G.~Aad$^\textrm{\scriptsize 85}$,
B.~Abbott$^\textrm{\scriptsize 113}$,
J.~Abdallah$^\textrm{\scriptsize 151}$,
O.~Abdinov$^\textrm{\scriptsize 11}$,
R.~Aben$^\textrm{\scriptsize 107}$,
M.~Abolins$^\textrm{\scriptsize 90}$,
O.S.~AbouZeid$^\textrm{\scriptsize 158}$,
H.~Abramowicz$^\textrm{\scriptsize 153}$,
H.~Abreu$^\textrm{\scriptsize 152}$,
R.~Abreu$^\textrm{\scriptsize 116}$
% ...
\bigskip
\\
% ...
$^{11}$ Institute of Physics, Azerbaijan Academy of Sciences, Baku, Azerbaijan\\
% ...
$^{85}$ CPPM, Aix-Marseille Universit{\'e} and CNRS/IN2P3, Marseille, France\\
% ...
$^{90}$ Department of Physics and Astronomy, Michigan State University, East Lansing MI, United States of America\\
% ...
$^{107}$ Nikhef National Institute for Subatomic Physics and University of Amsterdam, Amsterdam, Netherlands\\
% ...
$^{113}$ Homer L. Dodge Department of Physics and Astronomy, University of Oklahoma, Norman OK, United States of America\\
% ...
$^{116}$ Center for High Energy Physics, University of Oregon, Eugene OR, United States of America\\
% ...
$^{151}$ Institute of Physics, Academia Sinica, Taipei, Taiwan\\
$^{152}$ Department of Physics, Technion: Israel Institute of Technology, Haifa, Israel\\
$^{153}$ Raymond and Beverly Sackler School of Physics and Astronomy, Tel Aviv University, Tel Aviv, Israel\\
% ...
$^{158}$ Department of Physics, University of Toronto, Toronto ON, Canada\\
% ...
\end{flushleft}

%\end{document}
% Created with xml2latex.py

%-------------------------------------------------------------------------------
% Print the list of contributors to the analysis
% The argument gives the fraction of the text width used for the names
%-------------------------------------------------------------------------------
%\clearpage
%\PrintAtlasContribute{0.30}

%-------------------------------------------------------------------------------
% Auxiliary material
\ifauxmat
\clearpage
\appendix
\input{TOPQ_2015_01-auxmat}
\fi
%-------------------------------------------------------------------------------
%
%In an ATLAS paper, auxiliary plots and tables that are supposed to be made public 
%should be collected in an appendix that has the title \enquote{Auxiliary material}.
%This appendix should be printed after the Bibliography.
%At the end of the paper approval procedure, this information can be split into a separate document
%-- see \texttt{atlas-auxmat.tex}.
%
%In an ATLAS note, use the appendices to include all the technical details of your work
%that are relevant for the ATLAS Collaboration only (e.g.\ dataset details, software release used).
%This information should be printed after the Bibliography.

%\end{multicols}

\end{document}